\documentclass{PoS}

\title{Is Chemical Equilibrium achieved in Collisions of Small Systems at the SPS?}

\ShortTitle{Is Chemical Equilibrium achieved in Collisions of Small Systems at the SPS?}

\author{\speaker{Ingrid~Kraus} ~and Helmut~Oeschler\\
        Institut f\"ur Kernphysik, Technische Universit\"at Darmstadt, Germany\\
        E-mail: \email{Ingrid.Kraus@cern.ch}, \email{H.Oeschler@gsi.de}}

\author{Krzysztof Redlich\\
	Institute of Theoretical Physics, University of Wroc\l aw, Poland\\
        E-mail: \email{Redlich@ift.uni.wroc.pl}}

\abstract{
Ultrarelativistic nucleus-nucleus collisions are investigated with the goal to study the properties of strongly interacting matter under extreme conditions of high energy density.
Hadron multiplicities can provide information on the nature of the medium from which they are originating. The Statistical Model was recognised as a powerful approach to describe particle yields established at chemical decoupling. In general, this model assumes that at freeze-out the collision fireball appears as a statistical  system in thermal and  chemical equilibrium. The ensembles are constrained by charge conservation laws. 
\\
The Statistical Model has to be formulated in the canonical ensemble with respect to strangeness conservation if the number of strange particles becomes small. However, the canonical suppression under the assumption of strangeness chemical equilibrium in the whole fireball volume was found to be not sufficient to reproduce observed yields. Two approaches have been proposed to modify the model. First, a non-equilibrium factor $\gamma_S$ was introduced in canonical and grand-canonical ensembles as the additional fit parameter to account for the suppressed strange particle phase-space. Here we focus on the second method: The model is extended by correlation volumes which restrict the strangeness chemical equilibrium only to certain subvolumes of the system.
\\
In this work we report on the analysis of experimental data on particle production from p+p and central C+C, Si+Si and Pb+Pb collisions at the top SPS energy within the Statistical Model. The abundances of strange particles, in particular in the small systems, are found to be below the expectation of the Statistical Model formulated in the canonical ensemble. Therefore, we introduce strangeness equilibrated subvolumes. The canonical strangeness suppression in these correlated clusters accounts successfully for the smaller production of strange particles. The system size dependence of the correlation volume and of the thermal parameters are presented.
}

\FullConference{International Europhysics Conference on High Energy Physics\\
        July 21st - 27th 2005\\
        Lisboa, Portugal}

\begin{document}

\section{Introduction}
\begin{figure}
\vspace{-0.8cm}
\noindent
\begin{minipage}[b]{0.33\linewidth}
\centering
\includegraphics[width=\linewidth]{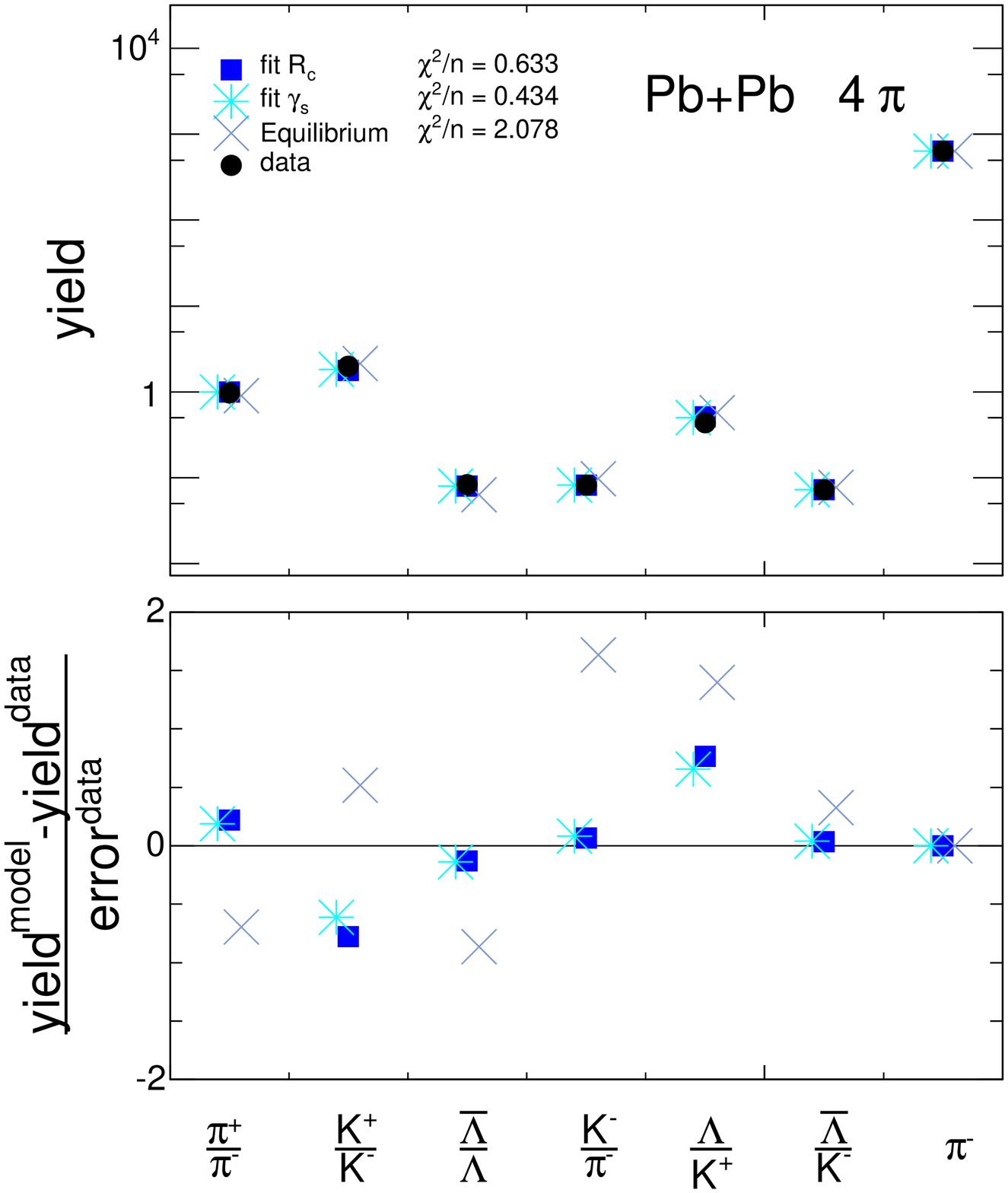}
\end{minipage}\hfill
\begin{minipage}[b]{0.33\linewidth}
\centering
\includegraphics[width=\linewidth]{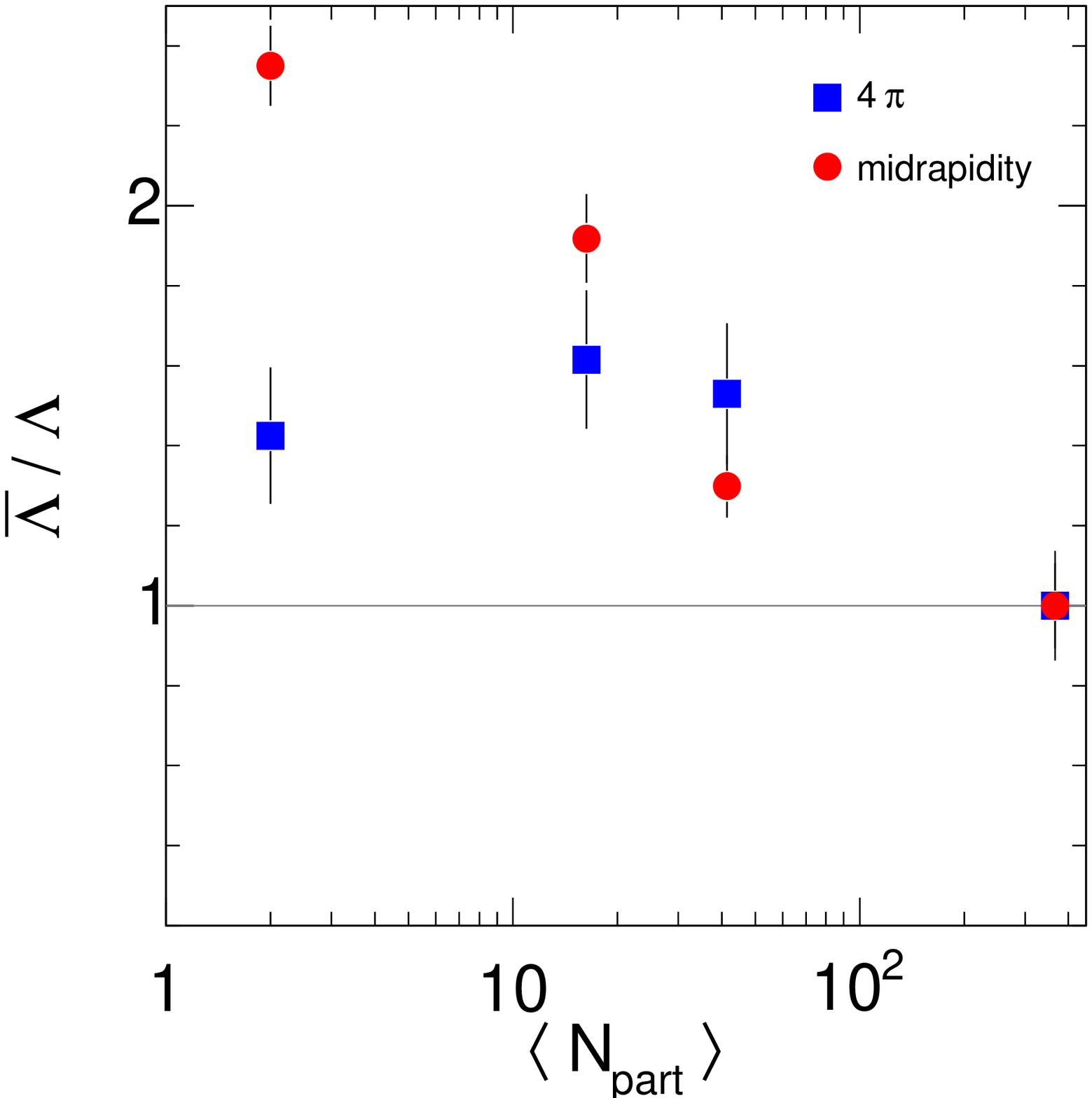}
\end{minipage}\hfill
\begin{minipage}[b]{0.33\linewidth}
\centering
\includegraphics[width=\linewidth]{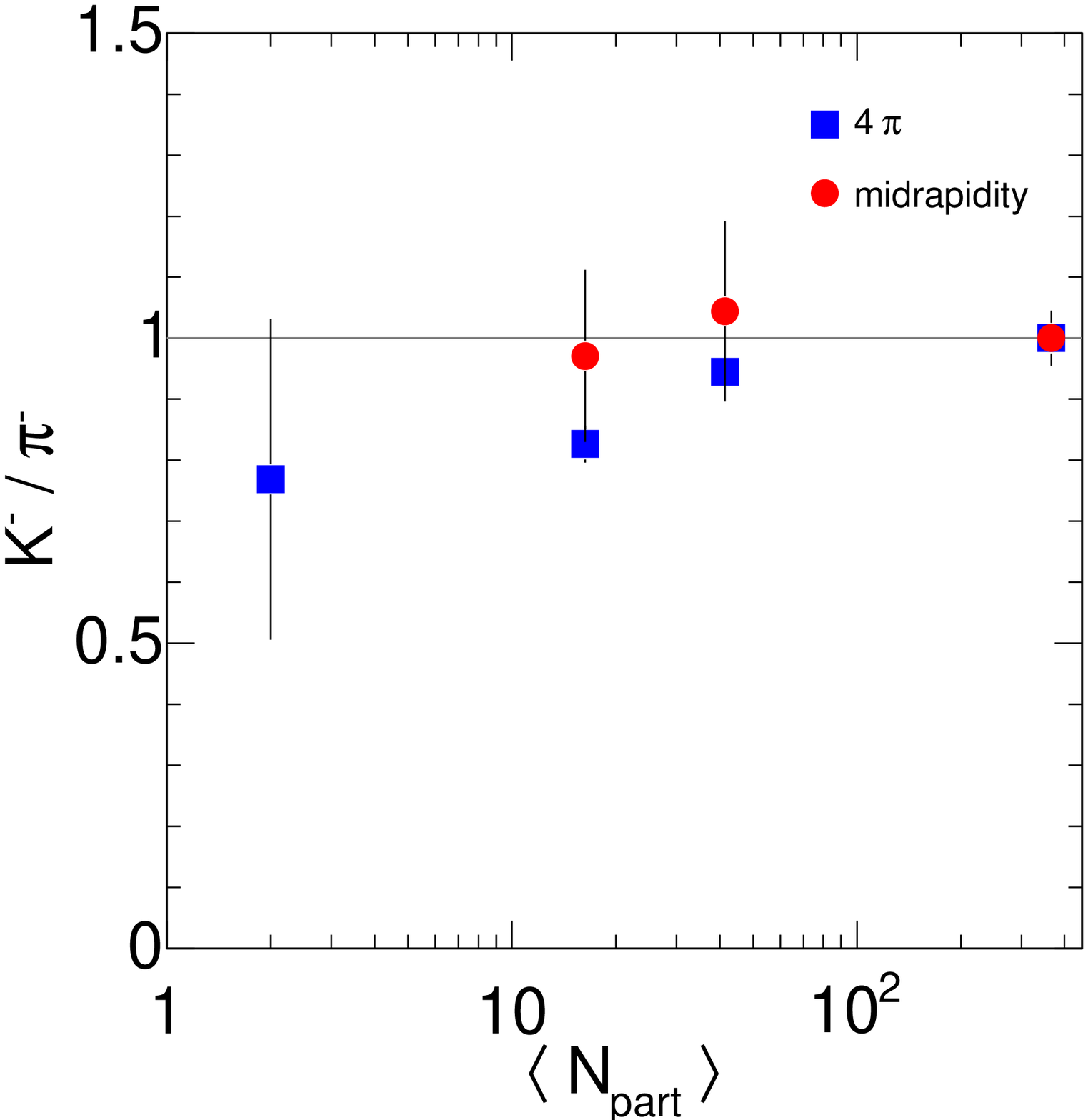}
\end{minipage}
\caption{Left: Integrated $\pi^-$ yield and particle ratios from Pb+Pb collisions at $\sqrt{s_{NN}}$~=~17.3~$A$GeV together with model fits (a) (crosses), (b)  (stars) and 		(c) (squares). The lower panel of the left side shows the deviation of the model fits to data.
		Middle and right: Midrapidity (squares) and 4$\pi$ (circles) $\bar{\Lambda}/\Lambda$ ratio (middle) and K$^-/\pi^-$ ratio (right) in p+p and central C+C, Si+Si and Pb+Pb collisions, normalised to the Pb+Pb measurement.
		}
\label{fig1} 
\end{figure}
Considering particle  production in heavy ion collisions, a particular role has been attributed to strange particles because strangeness was predicted to be a sensitive probe of the properties of QCD matter. The Statistical Model \cite{hwa} is very successful in describing the chemical composition of the final state of collisions over a wide range of incident energies. However, without an additional strangeness undersaturation factor, $\gamma_S$, Hadron Gas Models hardly reproduce the data from small colliding systems nor from reactions at the smaller collision energies. Here we investigate the influence of an alternative assumption, described in Section 2, on the model predictions.
\\
In this work we investigate the system size dependence of strangeness production within Statistical Model. Experimental results at the top SPS energy, $\sqrt{s_{NN}}$~=~17.3~$A$GeV, from p+p and central C+C, Si+Si and Pb+Pb collisions are studied~\cite{data}. Since only a limited number of hadrons were analysed in the small systems, we restricted our study on a consistent set of data. The investigated yield and ratios are demonstrated in Fig.~\ref{fig1} on the left hand side. Both, midrapidity densities and integrated yields were considered. Only the missing midrapidity data from p+p interactions make an exception.
\section{The model description and the basic assumptions}
The Statistical Model is applicable if the particles observed in high energy collisions are originating from a thermal fireball. This fireball of volume $V$ is characterised by thermal parameters,  the temperature $T$ and baryon chemical potential $\mu_B$, which are the same all over the system. Furthermore, we assume that strangeness is strongly correlated and appears in chemical equilibrium only within subvolumes $V_C$ of the system. Consequently there are two volume parameters in the model. The overall volume of the system, which determines the particle yields at  fixed density, and the correlation (cluster) volume $V_C$, which enters through the canonical suppression factor and reduces the densities of strange particles. Assuming spherical geometry the volume $V_C$ is parameterised by the radius $R_C$ which serves as a free parameter and defines the range of local equilibrium. A particle with strangeness quantum number $N_S$ can appear anywhere in the volume $V$, however it has to be accompanied within the subvolume $V_C$ by other particles carriing strangeness $-N_S$ to conserve strangeness exactly.
\\
Three different model settings are compared: (a) an equilibrium ansatz, which uses only the temperature $T$ and the baryon chemical potential $\mu_B$  to describe the data, (b) additionally the strangeness undersaturation factor $\gamma_S$ is introduced in the standard way (multiplicative factor for each (anti)strange valence quark), (c) subvolumes with correlated strangeness production are generated by the cluster radius $R_C$. All Statistical Model results presented in these proceedings were achieved with the THERMUS package~\cite{thermus}.
\begin{figure}
\vspace{-0.85cm}
\noindent
\begin{minipage}[b]{0.7\linewidth}
\centering
\includegraphics[width=\linewidth]{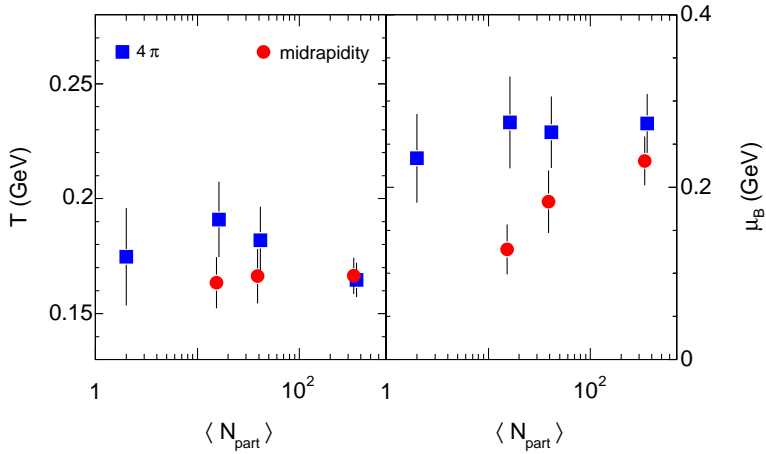}
\end{minipage}
\hfill
\begin{minipage}[b]{0.29\linewidth}
\centering
~ \\
\caption{Chemical freeze out temperature $T$ (left) and baryon chemical potential $\mu_B$ (right) from fits to midrapidity densities (circles) and integrated yields (squares)  from p+p and central C+C, Si+Si and Pb+Pb collisions, derived with model (c). Please note the suppressed zero in the left panel.
}
\vspace{1.75cm}
\end{minipage}
\vspace{-10mm}
\label{fig2} 
\end{figure}

\section{Results}
\begin{figure}
\vspace{-0.85cm}
\noindent
\begin{minipage}[b]{0.45\linewidth}
\centering
\includegraphics[width=\linewidth]{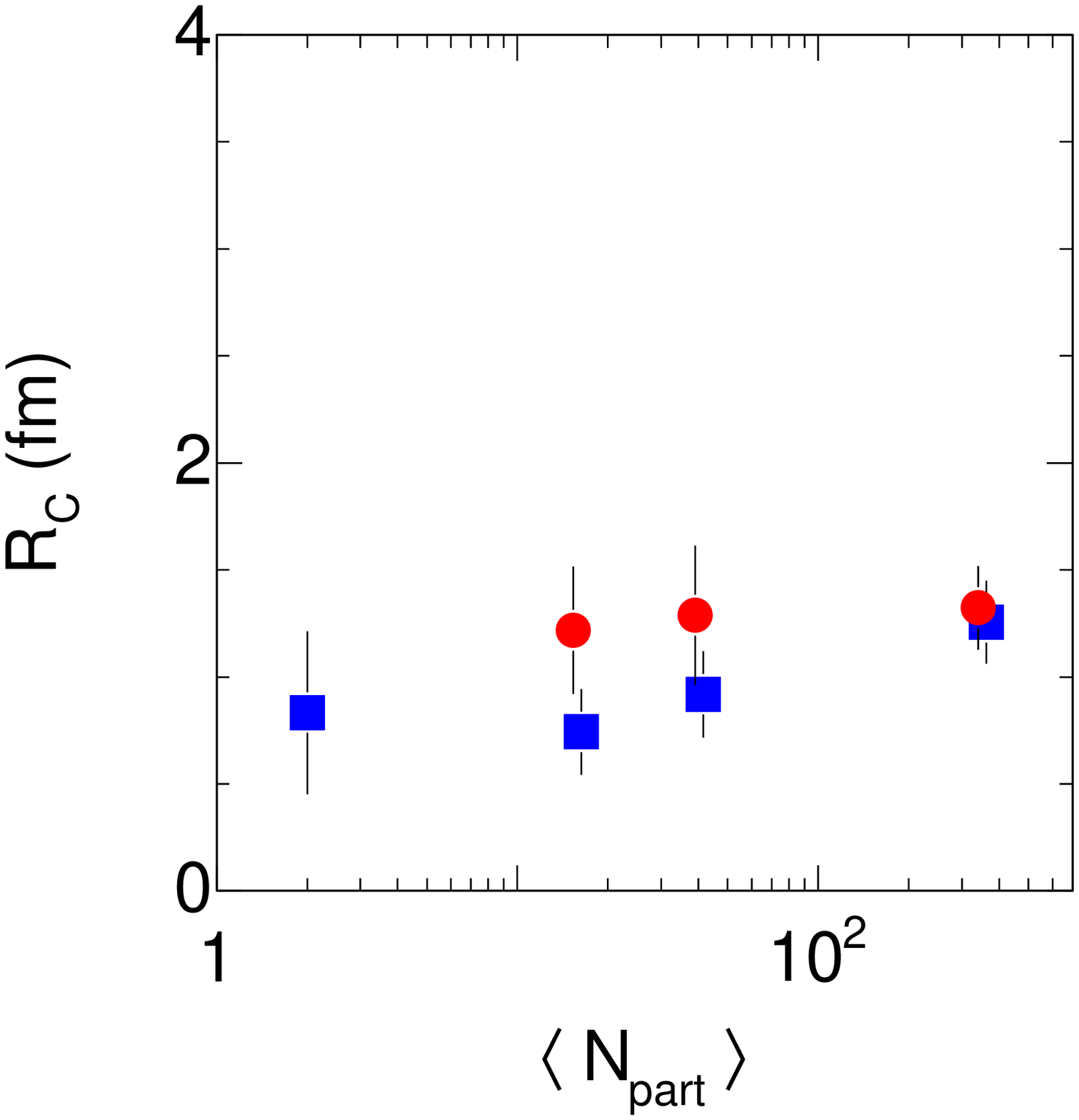}
\end{minipage}\hfill
\begin{minipage}[b]{0.45\linewidth}
\centering
\includegraphics[width=\linewidth]{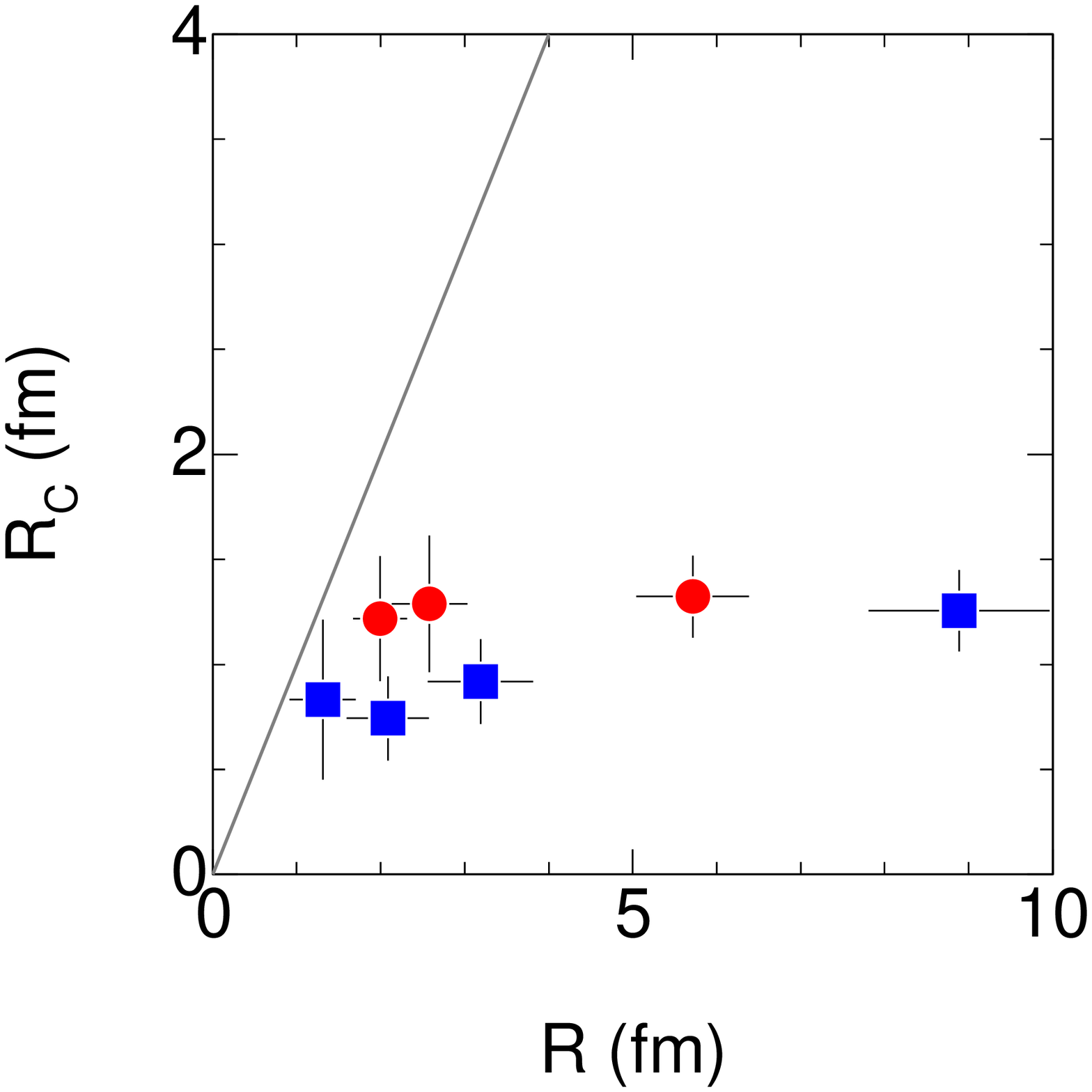}
\end{minipage}
\vspace{-5mm}
\caption{Cluster radius $R_C$ as a function of the number of participants (left) and as a function of the fireball radius $R$ (right) from fits to midrapidity densities (circles) and integrated yields (squares) from p+p and central C+C, Si+Si and Pb+Pb collisions. The line in the right panel indicates $R_C$ = $R$.
}
\label{fig3} 
\end{figure}

The results of the three model settings together with Pb+Pb data are shown on the left hand side of Fig.~\ref{fig1}. While the option (a) with completely equilibrated strangeness abundance fails (in particular for the small systems, not shown here), both, the approach with $\gamma_S$ and the one with $R_C$ yield comparable good descriptions of the data. These two models result in freeze-out temperatures $T$ which are barely dependent on the system size (see left hand side of Fig.~\ref{fig2} for model (c)). The baryon chemical potential is flat as a function of system size if total yields are considered. In contrast, $\mu_B$ is decreasing towards smaller systems in fits to midrapidity densities (Fig.~\ref{fig2} left hand side). This directly reflects the $\bar{\Lambda}/\Lambda$ ratio which indicates stronger stopping in larger systems (Fig.~\ref{fig1}, middle). ~
\\
The cluster radius, R$_C$, varies between 0.7~fm and 1.4~fm in all data sets under study (left hand side of Fig.~\ref{fig3}). Larger radii in the fits to midrapidity data are correlated with the experimental observation of larger K/$\pi$ ratios compared to 4$\pi$ yields (Fig.~\ref{fig1}, right hand side). The increase of strange to non-strange particle ratios by a factor of 2 from p+p to Pb+Pb reactions is reflected in a variation of the subvolume radius from about 0.7 to 1.4 fm.
\\
In the data set considered here, the fireball radius $R$ at freeze-out is determined by the multiplicity of $\pi^-$ mesons. The smaller midrapidity densities cause smaller radii $R$ than the integrated yields. In the comparison of the cluster to the fireball radius it becomes clear that R$_C$ has a significantly smaller system size dependence than $R$ (right hand side of Fig.~\ref{fig3}). 
In the smallest system, this correlation length is almost as large as the fireball radius. In contrast, with increasing system size, the subvolumes with locally conserved strangeness are up to six times smaller than the entire fireballs. In this respect, the fact that $R_C$ is of the order of 1~fm and only weakly increasing with system size is an astonishing result and requires further studies. From this analysis we conclude that the transition from suppressed to saturated strangeness production happens in a very narrow range of the correlation length $R_C$. This behavior is also known from percolation model calculations~\cite{satz}. 

To summarise, the arbitrary undersaturation factor $\gamma_S$ in the Statistical Model can be replaced by the concept of strangeness equilibrated subvolumes which through canonical suppression reproduces successfully the experimental data.

\end{document}